\documentclass[12pt]{article}

\usepackage{tikz}
\usetikzlibrary{decorations.pathreplacing,intersections}
\usetikzlibrary{calc}
\usepackage[margin=1in]{geometry} 
\usepackage{amsmath,amsthm,amssymb}
\usepackage{tikz}
\usepackage{natbib}
\usepackage{subcaption}
\usepackage[utf8]{inputenc}
\usepackage{enumitem}
\usepackage{setspace}
\usepackage{color}
\usepackage{soul}

\newtheorem{cl}{Claim}
\newtheorem{prop}{Proposition}[section]
\newtheorem{thm}{Theorem}[section]
\newtheorem{cor}{Corollary}[section]

\usepackage{cancel}
\newtheorem{lem}{Lemma}[section]
\theoremstyle{definition}
\newtheorem{dfn}{Definition}[section]
\newtheorem{algo}{Algorithm}[section]
\newtheorem{ex}{Example}[section]

\usepackage{ragged2e}
\definecolor{amber}{rgb}{1.0,0.75,0.0}
\definecolor{aqua}{rgb}{0,1,1}
\definecolor{amaranth}{rgb}{1,.6,.62}

\title{The Limits of Identification in Discrete Choice \footnote{We are thankful to two anonymous referees, the advisory editor, Roy Allen, and Yusufcan Masatlioglu for helpful comments and discussions during the course of this project. \\
Chambers: Department of Economics, Georgetown University.  E-mail:  \texttt{cc1950@georgetown.edu}\\
Turansick: Department of Decision Sciences and IGIER, Universit\'{a} Bocconi.  E-mail:  \texttt{christopher.turansick@unibocconi.it}}}

\author{Christopher P. Chambers and Christopher Turansick}
\date{\today}

\begin{document}

\maketitle

\begin{abstract}
This paper uncovers tight bounds on the number of preferences permissible in identified random utility models.  We show that as the number of alternatives in a discrete choice model becomes large, the fraction of preferences admissible in an identified model rapidly tends to zero.  We propose a novel sufficient condition ensuring identification, which is strictly weaker than some of those existing in the literature. While this sufficient condition reaches our upper bound, an example demonstrates that this condition is not necessary for identification. Using our new condition, we show that the classic ``Latin Square" example from social choice theory is identified from stochastic choice data.
\end{abstract}

\section{Introduction}
The random utility model is the basis for much of modern discrete choice analysis. Random utility moves beyond the classic rational model and supposes that there is some distribution over preferences inducing choice. This distribution over preferences is typically interpreted as heterogeneity within a population or variation of a single agent's preference across time. A common desideratum for a model in empirical practice is identification.  Identification allows for proper counterfactual and welfare analysis. The random utility model is well-known to be unidentified, however there are many refinements of the random utility model which recover identification. We study those models which recover identification by limiting the set of allowable preferences in the model.

Identifying assumptions by nature put restrictions on a model in order to recover identification.  We study identifying restrictions which assert that preferences must belong to some prespecified class.  Our main result shows that the maximum number of preferences allowed for a set of preferences leading to an identified random utility model is equal to $(n-2)2^{n-1}+2$, where $n$ is the number of alternatives in our environment. It follows that the ratio of preferences in such a model to the total number of preferences given a set of alternatives of size $n$ scales at a rate of $\mathcal{O}(\frac{2^{n-1}}{(n-1)!})$. This tells us that, as $n$ grows large, the number of preferences in a model identified through preference restrictions is vanishingly small when compared to the total number of possible preferences. When $n=5$, any identified model must rule out at least half of the possible preferences.  When $n=9$, more than $99.5\%$ of the preferences must be discarded in order to write an identified model. Thus identifying assumptions of this nature must assume away almost all types of behavior.

Our second goal in this paper is to offer a new identifying restriction for random utility models. Our new identifying restriction is called \textit{edge decomposability}. In our setup a random utility model is simply a collection of feasible preferences. Edge decomposability asks that, for a given model, for every non-empty subset of preferences of that model, there exists some preference in that subset with an upper contour set unique to that preference within that subset. More specifically it asks that for every non-empty subset of preferences, there exists some pair $(x,A)$, where $x$ is an alternative in menu $A$, such that there is a single preference in this subset that chooses $x$ from $A$ but fails to choose $x$ from any strict superset of $A$. Edge decomposability is a sufficinet condition for identification of a model. Among identified models, edge decomposability has large explanatory power. This is due to the fact that the largest edge decomposable model is the same size as the largest identified model. However, we are able to show through an example that there are identified models which are not edge decomposable.

We then compare edge decomposability with the identifying restrictions of \citet{apesteguia2017single} and \citet{turansick2022identification}. We show that every set of preferences which can be the support of some distribution over preferences satisfying the conditions of either \citet{apesteguia2017single} or \citet{turansick2022identification} must also be edge decomposable. Further, we show that there are sets of preferences that are edge decomposable yet cannot be the support of a distribution satisfying either the conditions of \citet{apesteguia2017single} or \citet{turansick2022identification}.  Finally, we introduce a new model of random utility which captures the classic Latin square example from social choice. The Latin square example is a collection of preferences that lead to Condorcet cycles when aggregating preferences via majority rule. We use our edge decomposability condition to show that this model is identified from stochastic choice data.

The rest of this paper proceeds as follows. In Section~\ref{model} we introduce our setup and some preliminary constructions. In Section~\ref{Dimension} we provide our main result and discuss its proof. In Section~\ref{EdgeDecSection} we introduce our identifying restriction and compare it to the restrictions of \citet{apesteguia2017single}, \citet{turansick2022identification}, as well as the Latin square example. Finally, we conclude with a discussion of the related literature in Section \ref{RelLit}.

\section{Model and Preliminaries}\label{model}

We work with a finite set of alternatives $X_n=\{1,\dots,n\}$ where each alternative is indexed by a number between $1$ and $n$. Further, we use $x$ and $y$ to denote arbitrary elements of $X_n$. We are interested in strict preferences over this environment.  Denote the set of linear orders over $X_n$ by $\mathcal{L}(X_n)$.\footnote{A linear order is complete, antisymmetric, and transitive.} A typical linear order is denoted by $\succ$.
\begin{dfn}
    A \textbf{model on $X_n$} (or simply a model) is a nonempty set of linear orders $M$:  $\emptyset \subsetneq M \subseteq \mathcal{L}(X_n)$.
\end{dfn}
We denote the collection of all models on $X_n$ as $\mathcal{M}_n$. Our notion of model does not capture every possible type of restriction. In our setup, a model restricts which preferences an agent may have. Other random utility models may be constructed through parametric restrictions.

\begin{ex}[Social Choice and Latin Squares]\label{latinsquare1}
    Suppose there is some exogenous ordering $\rhd$ over alternatives $X_n$. Without loss we can let it be $1 \rhd 2 \rhd \dots \rhd n$. We can consider the model $M$ which forms the Latin square given this ordering. In this model $M$, there is one preference $\succ_m$ for each $m\in \{1,\dots,n\}$. $\succ_m$ is given by $m \succ m+1 \succ \dots \succ n-1 \succ n \succ 1 \succ \dots \succ m-1$. We then consider distributions over this model $M$. This example considers a restriction on the set of available preferences, and thus is a model in our setup.
\end{ex}

\begin{ex}[Luce Random Choice]\label{luce}
    Consider the following class of random choice rules, described by \citet{luce1959individual}. Each $x \in X$ is given a weight $w(x) \in (0,\infty)$. The probability that $x$ is chosen from a menu $A$ is then given by $\frac{w(x)}{\sum_{y\in A} w(y)}$. This class of choice rules is often called the ``Luce model,'' though it is not a model in our sense.  Rather, instead of being parametrized by a distribution over preferences, the parameters in this model correspond to $w(x)$.  These parameters are not uniquely identified, but the normalized version $\frac{w(x)}{\sum_{y\in X}w(y)}$ are. 
\end{ex}
Using the standard $\Delta$ notation to represent the set of probability measures over a set, let $\nu \in \Delta(M)$ denote a typical probability distribution over the preferences in model $M$. Further, let $2^X$ denote the power set of $X$ and let $A$ denote a typical element of $2^X$.
\begin{dfn}
    A function $p:X_n \times 2^{X_n} \setminus \{\emptyset\} \rightarrow \mathbb{R}$ is a \textbf{random choice rule} if it satisfies the following.
    \begin{itemize}
        \item $p(x,A) \geq 0$ for all $x \in A$
        \item $\sum_{x \in A}p(x,A)=1$
    \end{itemize}
\end{dfn}
Random choice rules correspond to our data primitive. They denote how frequently an alternative $x$ is chosen from a menu $A$. Given $\nu\in\Delta(\mathcal{L}(X_n))$, let $p_\nu$ denote the random choice rule induced by $\nu$. Specifically,\footnote{Here, $\mathbf{1}\{ \cdot \}$ corresponds to the standard indicator function. It returns $1$ if the logical statement in the brackets is true and returns $0$ otherwise.} 
\begin{equation}
    p_\nu(x,A)=\sum_{\succ\in \mathcal{L}(X_n)} \nu(\succ)\mathbf{1}\{x \succ y \text{ }\forall y \in A \setminus \{x\}\}.
\end{equation}
\begin{dfn}
    A model $M$ is \textbf{identified} if for any $\nu, \nu' \in \Delta(M)$, $p_\nu=p_{\nu'}$ implies that $\nu=\nu'$.
\end{dfn}
Given $X_n$, we denote the set of identified models as $\mathcal{I}_n$. Our goal is to provide a tight upper bound on the size of models in $\mathcal{I}_n$. Specifically, our main question is what is the largest number of preferences that admit an identified model. In order to answer this question, we rely on the M\"{o}bius inverse of a random choice rule.\footnote{The M\"{o}bius inverse is a fairly general method of inverting cumulative sums on certain partially ordered sets.  In our case, the partially ordered set under consideration is $(X_n, \subseteq)$, and the inversion formula is referred to as the inclusion-exclusion principle.  A classical reference is \citet{rota1964foundations}.}
\begin{dfn}
    The \textbf{M\"{o}bius inverse} of $p(x,A)$ is a function $q:X \times 2^{X}\setminus \{\emptyset\} \rightarrow \mathbb{R}$ defined as follows.
    \begin{equation}\label{mobinvform}
        \begin{split}
            q(x,A) & = p(x,A) - \sum_{A \subsetneq B}q(x,B) \\
            & = \sum_{A \subseteq B} (-1)^{|B\setminus A|}p(x,B)
        \end{split}
    \end{equation}
\end{dfn}
The M\"{o}bius inverse $q(x,A)$ simply keeps track of how much choice probability is being added to or removed from the choice of $x$ at menu $A$. Further, the M\"{o}bius inverse function has a special relationship with random choice rules induced by models.
\begin{dfn}
    A random choice rule $p$ is \textbf{stochastically rational} if there exists some $\nu \in \Delta(\mathcal{L}(X))$ such that $p=p_\nu$.
\end{dfn}
Consider the set of preferences which choose $x$ from $A$ but do not choose $x$ from any superset of $A$. This set is given by $L(x,A)=\{\succ| z \succ x \succ y \text{ }\forall y\in A\setminus \{x\}, \text{ }\forall z \in X \setminus A\}$.
\begin{thm}[\citet{falmagne1978representation}]\label{falmagneident}
Let $q$ be the M\"{o}bius inverse of $p_{\nu}$.  Then for all $A\subseteq X$ and all $x\in A$, $q(x,A)=\nu(L(x,A))$.
\end{thm}

This result tells us that $q(x,A)$ is the probability under $p_{\nu}$ of $L(x,A)$. Further, our main result relies on a graphical representation of random choice rules. This directed graph is defined as follows. There is one node of this graph for each $A \in 2^X$. We index these nodes via $2^X$ and use the set $A$ to refer to the node indexed by $A$. There is a directed edge connecting $A$ to $B$ if $B=A \setminus \{x\}$ for some $x \in A$. We sometimes use $e$ to refer to an arbitrary edge and $e(A,B)$ to refer to the edge connecting node $A$ to node $B$. Finally, we assign capacities to each edge. The edge capacity of the edge connecting $A$ and $A \setminus \{x\}$ is given by $q(x,A)$. We call this graphical representation the \emph{probability flow diagram}.\footnote{The probability flow diagram is originally due to \citet{fiorini2004short}.}

\begin{figure}
    \centering
\begin{tikzpicture}[scale=.5, transform shape]
    \tikzstyle{every node} = [rectangle]
    
        \node (a) at (6,0) {$\emptyset$};
        
        \node (b) at (0,5) {$\{x\}$};
        \node (c) at (6,5) {$\{y\}$};
        \node (d) at (12,5) {$\{z\}$};
        
        \node (e) at (0,10) {$\{x,y\} $};
        \node (f) at (6,10) {$\{x,z\}$};
        \node (g) at (12,10) {$\{y,z\}$};

        \node (h) at (6,15) {$\{x,y,z\}$};
        
        \draw [->] (h) -- (f) node[midway, below, sloped] {$q(y,\{x,y,z\})$};
        \draw [->] (h) -- (g) node[midway, below, sloped] {$q(x,\{x,y,z\})$};  
        \draw [->] (h) -- (e) node[midway, below, sloped] {$q(z,\{x,y,z\})$};

        \draw [->] (e) -- (b) node[midway, below, sloped] {$q(y,\{x,y\})$}; 
        \draw [->] (e) -- (c) node[pos=.25, below, sloped] {$q(x,\{x,y\})$};
        \draw [->] (f) -- (b) node[pos=.25, below, sloped] {$q(z,\{x,z\})$};
        \draw [->] (f) -- (d) node[pos=.25, below, sloped] {$q(x,\{x,z\})$};
        \draw [->] (g) -- (c) node[pos=.25, below, sloped] {$q(z,\{y,z\})$};
        \draw [->] (g) -- (d) node[midway, below, sloped] {$q(y,\{y,z\})$}; 

        \draw [->] (b) -- (a) node[midway, below, sloped] {$q(x,\{x\})$};
        \draw [->] (c) -- (a) node[midway, below, sloped] {$q(y,\{y\})$};
        \draw [->] (d) -- (a) node[midway, below, sloped] {$q(z,\{z\})$};
    
    \end{tikzpicture}
    \caption{The probability flow diagram for the set $X_n = \{x,y,z\}$.}
    \label{fig:probflow}
\end{figure}
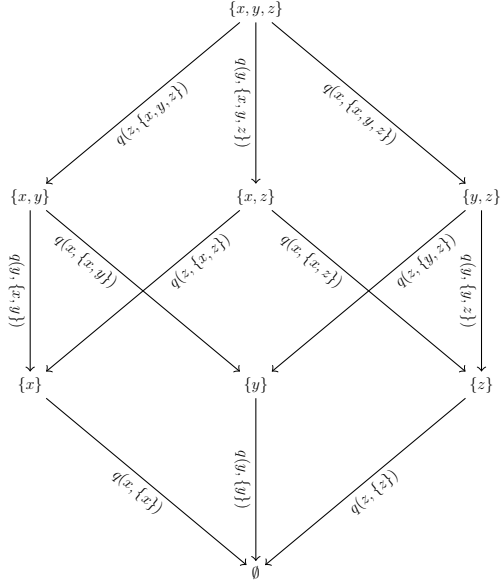

\begin{dfn}\label{pathdef}
A path $\rho$ is a finite sequence of sets $\{A_i\}_{i=0}^{n}$ such that $A_{i+1}\subsetneq A_i$ for all $i$, $A_0=X$, and $A_{|X|}=\emptyset$.
\end{dfn}
Observe that for every path, $|A_i \setminus A_{i+1}|=1$.  In the probability flow diagram, each path from $X_n$ to $\emptyset$ bijectively corresponds to a preference.  Each edge of the path corresponds to a unique input $(x,A)$ for the M\"{o}bius inverse. Using Theorem \ref{falmagneident}, a path that goes from $A_0$ to $A_n$ corresponds to the preference $A_0 \setminus A_1 \succ A_1 \setminus A_2 \succ \dots \succ A_{n-1}\setminus A_n$. To further describe this connection, if a random choice rule $p$ is induced by choice according to $\succ$, then each $q(x,A)$ on the path corresponding to $\succ$ is equal to one and each $q(y,B)$ not on the path corresponding to $\succ$ is equal to zero.

\section{The Dimension of Preferences}\label{Dimension}
 In this section we present our main result and proof. 

 \begin{thm}\label{prefbound}
     For any $n$, $\max_{M \in \mathcal{I}_n}|M|=(n-2)2^{n-1}+2$.
 \end{thm}

 Theorem \ref{prefbound} tells us that, if we have an identified model, then there are at most $(n-2)2^{n-1}+2$ preferences in that model. There are models which are identified and have fewer preferences, unidentified and have fewer preferences, and there are models which face intuitive restrictions but are not identified.

 \begin{ex}[Social Choice and Latin Squares Revisited]\label{latinsquare2}
     Recall the setup of Example \ref{latinsquare1}. If our model forms a Latin square then there are $n\leq (n-2)2^{n-1}+2$ preferences in the model, with a strict inequality when $n \geq 3$. We will show later that models that form Latin squares are identified.
 \end{ex}

 \begin{ex}[Single-Crossing Random Utility]\label{scrum1}
 Suppose there is some exogenous ordering $\rhd$ over alternatives $X_n$ for $n \geq 2$. Without loss we can let it be $1 \rhd 2 \rhd \dots \rhd n$. An ordered set of preferences $\{\succ_i\}_{i=1}^m$ satisfies the single-crossing property if $x \rhd y$, $x \succ_i y$, and $j \geq i$ imply $x \succ_j y$. \citet{apesteguia2017single} consider random utility models with single-crossing supports. A consequence of their results is that a model satisfying the single crossing property is identified. The maximal size of a model satisfying the single-crossing property is $\binom{n}{2}+1 \leq (n-2)2^{n-1}+2$ with a strict inequality when $n \geq 3$.

 \begin{prop}\label{scrumSize}
     Given a set of alternatives $X_n$ for $n \geq 2$, the maximum size of a set of preferences satisfying the single-crossing property is $\binom{n}{2}+1$.
 \end{prop}
     
 \end{ex}

\begin{ex}[Fishburn and Identification Failure]\label{FishNum}
The classic counterexample to identification of the random utility model is due to \citet{fishburn1998stochastic}. Let $X_n = \{a,b,c,d\}$. Consider the following probability distributions over linear orders on $X$.
\begin{equation*}
    \nu_1(\succ) =    \begin{cases}
                    \frac{1}{2} & \text{if }\succ \in \{ a \succ b \succ c \succ d, b \succ a \succ d \succ c\} \\
                    0 & \text{otherwise}
                 \end{cases}
\end{equation*}
\begin{equation*}
    \nu_2(\succ) =    \begin{cases}
                    \frac{1}{2} & \text{if }\succ \in \{ a \succ b \succ d \succ c, b \succ a \succ c \succ d\} \\
                    0 & \text{otherwise}
                 \end{cases}
\end{equation*}
These two probability distributions induce the same random choice rule. If we consider the model given by the four preferences in this example, we have a model which is not identified. The size of this model is $4 < (4-2)2^{4-1}+2=18$.
     
\end{ex}

\begin{ex}[Minimal Mutual Agreement]
Consider a model $M$. We say that a model satisfies \textit{minimal mutual agreement} if for each $\succ,\succ' \in M$, there are $x,y \in X_n$ such that $x \succ y$ and $x \succ' y$. Minimal mutual agreement may be assumed in cases where an analyst is trying to aggregate preferences. When minimal mutual agreement fails, there is a pair $\succ$, $\succ'$ of preferences that are directly opposed, and then any gain by $\succ$ must be accompanied by a loss in $\succ'$.  Given $\succ$, define $\succ^-$ by $x \succ y \implies y \succ' x$. It is easy to see that $M$ satisfies minimal mutual agreement if and only if ($\succ \in M \implies \succ^- \not \in M$) holds. It then follows that the maximal size of a model satisfying minimal mutual agreement is $\frac{n!}{2} > (n-2)2^{n-1}+2$ (for large enough $n$).
    
\end{ex}

Before presenting our proof of Theorem~\ref{prefbound}, we first discuss some of its implications. Recall that for a given $X_n$, there are $n!$ strict preferences. We now compare the number of possible preferences  with the maximal number of preferences in an identified model.  Since preferences here are linear orders, the number of possible preferences is $n!$.  According to Theorem~\ref{prefbound}, for reasonably large $n$, the size of a maximal identified model roughly doubles when adding a new alternative, while the size of the largest possible model, identified or otherwise, increases by a factor of $n$.  In terms of ratios,

\begin{equation}
    \begin{split}
        \frac{(n-2)2^{n-1}+2}{n!} & = \frac{2^{n-1}}{n(n-1)(n-3)!}+\frac{2}{n!} \\
        & \leq \frac{2^{n-1}}{(n-1)!}+\frac{2}{n!} \\
        & \leq \frac{2^{n-1}}{(n-1)!}+\frac{2}{(n-1)!} \\
        & = \frac{2^{n-1}+2}{(n-1)!}.
    \end{split}
\end{equation}

This tells us that the ratio of the maximal number of preferences in an identified model and the total number of preferences grows at a rate $\mathcal{O}(\frac{2^{n-1}}{(n-1)!})$. Clearly, as $n$ gets large, this ratio rapidly goes to zero.  Thus, any identified model must rule out a large mass of preferences.

We now discuss the proof of Theorem \ref{prefbound}. First, observe that there is an important connection between identified models and linear independence.

\begin{dfn}
    A set of vectors $\{\mathbf{v}_1,\dots,\mathbf{v}_n\}$ is \textbf{linearly independent} if any set of scalars $\{c_1,\dots,c_n\}$ with $\sum_{i=1}^n c_i \mathbf{v}_i=0$ implies that $c_i=0$ for all $i$.
\end{dfn}

Let $p_{\succ}$ denote the real-valued vector, indexed by elements of the form $(x,A)$ with $x\in A$, which encodes the random choice rule induced by choice according to $\succ$. In other words, $p_{\succ}(x,A)=1$ if $x \succ y$ for all $y \in A\setminus \{x\}$ and $p_{\succ}(x,A)=0$ otherwise. 

The following result is straightforward and known, but we give a proof of it in the appendix for completeness.

\begin{lem}\label{Plinind}
A model $M$ is identified if and only if the set $\{p_\succ\}_{\succ \in M}$ is linearly independent.\end{lem}

Now let $q_{\succ}$ denote the vector, indexed by elements of the form $(x,A)$ with $x\in A$, which encodes the M\"{o}bius inverse of the random choice rule induced by choice according to $\succ$. That is, $q_{\succ}(x,A)=1$ if $\succ \in L(x,A)$ and $q_{\succ}(x,A)=0$ otherwise. It is less obvious that a model $M$ is identified if and only if the set $\{q_\succ\}_{\succ \in M}$ is linearly independent.  It essentially follows from Lemma~\ref{Plinind} as the map carrying $p_{\succ}$ to $q_{\succ}$ is known to be linear and invertible. We provide a formal proof of the following result in the appendix.

\begin{lem}\label{qLinIndlem}
    A model $M$ is identified if and only if $\{q_\succ\}_{\succ \in M}$ is linearly independent.
\end{lem}

Our proof of Theorem \ref{prefbound} relies on the probability flow diagram with one small alteration. Specifically, we take the probability flow diagram and add an edge connecting $\emptyset$ to $X_n$ (and assign it an edge capacity equal to one). This motivates the following proposition.

\begin{prop}\label{add1prop}
Let $\{x^i\}_{i=1}^I$ be a collection of vectors for which for all $i,j$, $\sum_{k}x_k^i=\sum_{k}x_k^j\neq 0$.  Then $\{x^i\}_{i=1}^I$ is linearly independent if and only if $\{(x^i,1)\}_{i=1}^I$ is linearly independent, where $(x^i,1)$ refers to the concatenation of the vector $x^i$ with a $1$.
\end{prop}

In the proof of Theorem \ref{prefbound}, we make use of two concepts from graph theory. First, a circuit is a sequence of directed edges where one edge's starting node is the previous edge's ending node with the last edge's ending node being the same as the first edge's starting node. In simpler terms, a circuit is a directed cycle. Second, the cyclomatic number of a directed graph corresponds to the minimal number of edges in the graph needed to be removed in order to remove all cycles from the graph. The cyclomatic number also corresponds to the circuit rank of the graph or the maximum number of independent circuits in the graph. We are now ready to proceed with the proof of Theorem~\ref{prefbound}.

\begin{proof}
Terminology in this proof is as in \citet{berge2001theory}.    By Lemma~\ref{qLinIndlem} and Proposition~\ref{add1prop}, a model $M$ is identified if and only if the set of vectors $\{(q_{\succ},1)\}_{\succ\in M}$ is linearly independent.\footnote{Observe that for each $\succ$, $\sum_{(x,A):x\in A}q_{\succ}(x,A)=n$, so that the condition in Lemma~\ref{qLinIndlem} is satisfied.}

Now, consider the directed graph whose nodes are $2^{X_n}$ and has a directed edge from $A$ to $A\setminus \{x\}$ for any $x\in A$, and also has a directed edge from $\varnothing$ to $X_n$.  We call this the \emph{appended probability flow diagram}.  The vector $(q_{\succ},1)$ then represents an indicator function of a subset of the appended probability flow diagram.  The discussion following Definition~\ref{pathdef} explains that $(q_{\succ},1)$ represents an indicator function of a circuit on the appended probability flow diagram, or an oriented loop (the $1$ reflects the edge going from $\varnothing$ to $X_n$).  
    
By Theorem 3 of Chapter 4 of \citet{berge2001theory}, the cyclomatic number of the appended probability flow diagram is equal to the dimension of the space spanned by indicator functions of circuits of the appended probability flow diagram.  Now, the indicator function of any circuit of this diagram must be a linear combination of indicator functions of circuits corresponding to preferences.  The reasoning is straightforward:  by construction, every circuit must pass through the edge connecting $\varnothing$ to $X_n$.  A circuit which passes through this edge only once clearly corresponds to a preference.  A circuit passing through it $k$ times corresponds to a concatenation of $k$ circuits passing through this edge only once.  Hence, the indicator function of the circuit passing through this edge $k$ times is the sum of the indicator functions of the $k$ circuits which pass through the edge only once, each of which correspond to a preference. Hence, the dimension of the circuit space of the appended probability flow diagram is the same as the dimension of the space spanned by $\{(q_{\succ},1)\}$ for all preferences $\succ$.

The cyclomatic number of a strongly connected graph, as our appended probability flow diagram is, according to \citet{berge2001theory}, is defined by $E-N+1$, where $E$ is the number of edges and $N$ is the number of nodes. There are $2^{n}$ nodes in our appended probability flow diagram. There is one edge in our probability flow diagram for each $(x,A)$ with $x \in A$. This is given by $\sum_{i=1}^{n} i \binom{n}{i}$, which is equal to $n2^{n-1}$.\footnote{The argument is standard, for each element of a set of size $n$, there are $2^{n-1}$ sets containing it.  So $\sum_{i=1}^n i\binom{n}{i}$, which counts the sum of cardinalities of all subsets, must be given by summing the number $2^{n-1}$ across each element of the set of size $n$, resulting in $n2^{n-1}$.  See \citet{feller}, p.63.} This means that our appended probability flow diagram has $n2^{n-1}+1$ edges, accounting for the edge going from $\varnothing$ to $X_n$. It then follows that the cyclomatic number of our appended probability flow diagram is given by the following:
    \begin{equation}
        n2^{n-1}+1-2^n+1 = (n-2)2^{n-1}+2.
    \end{equation}

    Thus, the space spanned by $(q_{\succ},1)$ has dimension $(n-2)2^{n-1}+2$.  This tells us that for any model $M$, the dimension of the space spanned by $(q_{\succ},1)$ as $\succ\in M$ is at most $(n-2)2^{n-1}+2$, and that this bound is achieved for some model.  In particular, any $M$ for which $(q_{\succ},1)$ is linearly independent must have $|M|\leq (n-2)2^{n-1}+2$ and this bound is achieved for some $M$.  Again by Proposition~\ref{add1prop}, any $M$ for which $q_{\succ}$ is linearly independent must have $|M|\leq (n-2)2^{n-1}+2$ and this bound is achieved.  Conclude by Lemma~\ref{qLinIndlem} that any identified model must have $|M|\leq (n-2)2^{n-1}+2$ and that this bound is achieved. \end{proof}

The dimension of all random choice rules is the same as the dimension of the set of RUM random choice rules; see \emph{e.g.} Corollary 6(ii) of \citet{saito2018axiomatizations} or Theorem 2 of \citet{dogan2022every}, where it is shown that the rational choice functions span the span of all stochastic choice functions.  This immediately gives us the following corollary to Theorem~\ref{prefbound}, which is not so difficult to prove directly by simple combinatorial arguments.

\begin{cor}\label{cor:bound}The set of random choice rules on $X_n$ has dimension $(n-2)2^{n-1}+2$.\end{cor}

Corollary~\ref{cor:bound} obtains bounds on identified models even absent the RUM restriction.

\section{A Sufficient Condition for Linear Independence}\label{EdgeDecSection}
In this section, using the graphical intuition developed in the last section, we propose a new condition which guarantees that a model is identified. While this condition is sufficient for identification and reaches the upper bound from Theorem \ref{prefbound}, we show by example that it fails to be necessary. We then compare this condition to two other uniqueness conditions in the random utility literature and the Latin square construction from Example \ref{latinsquare1}. We now introduce our new condition.

\begin{dfn}
    We say that a model $M$ is \textbf{edge decomposable} if for every nonempty submodel $N \subseteq M$ there exists a preference $\succ \in N$ and a tuple $(x,A)$ with $x \in A$ such that $N \cap L(x,A)=\{\succ\}$.
\end{dfn}

Our arguments often leverage a form of edge decomposability which takes the form of an induction argument:

\begin{dfn}
    We say that a model $M$ is \textbf{sequentially edge decomposable} if it can be enumerated as $\{\succ_1,\ldots,\succ_k\}$ so that for each $i\in \{1,\ldots,k\}$, there is a tuple $(x,A)$ with $x\in A$ such that $\{\succ_i,\ldots,\succ_k\}\cap L(x,A)=\{\succ_i\}$.
\end{dfn}

A simple induction argument guarantees the following, which we will often use without mention.

\begin{prop}A model $M$ satisfies edge decomposabilitiy iff it satisfies sequential edge decomposability.\end{prop}    

Keeping in mind Theorem \ref{falmagneident}, when a model is edge decomposable, we are able to do the following to recover the unique distribution over preferences in our model which induces the data. First, as $M$ is a (weak) submodel of $M$, there exists some preference $\succ$ and pair $(x,A)$ such that $\{\succ\}=M \cap L(x,A)$. By Theorem \ref{falmagneident}, this tells us that $q(x,A)=\nu(L(x,A))=\nu(\succ)$. Now consider the submodel which is given by $N=M\setminus \{\succ\}$. By edge decomposability, we know that there is some $\succ' \in N$ such that $\{\succ'\}=N\cap L(y,B)$ for some $y \in B$. It then follows that $\nu(\succ')=q(y,B)$ if $\succ \not \in L(y,B)$ and $\nu(\succ')=q(y,B)-\nu(\succ)$ if $\succ \in L(y,B)$. Since $M$ is finite, we can repeated this argument to get full recovery of our distribution $\nu$ over preferences in $M$, and so every edge decomposable model $M$ is identified. The following theorem formally restates this observation and collects the identification properties of edge decomposability.

\begin{thm}\label{edgeIdentification}
    \begin{enumerate}
        \item If a model $M$ is edge decomposable, it is identified.
        \item The maximum size of an edge decomposable model is $(n-2)2^{n-1}+2$.
        \item There are identified models which are not edge decomposable.
    \end{enumerate}
\end{thm}

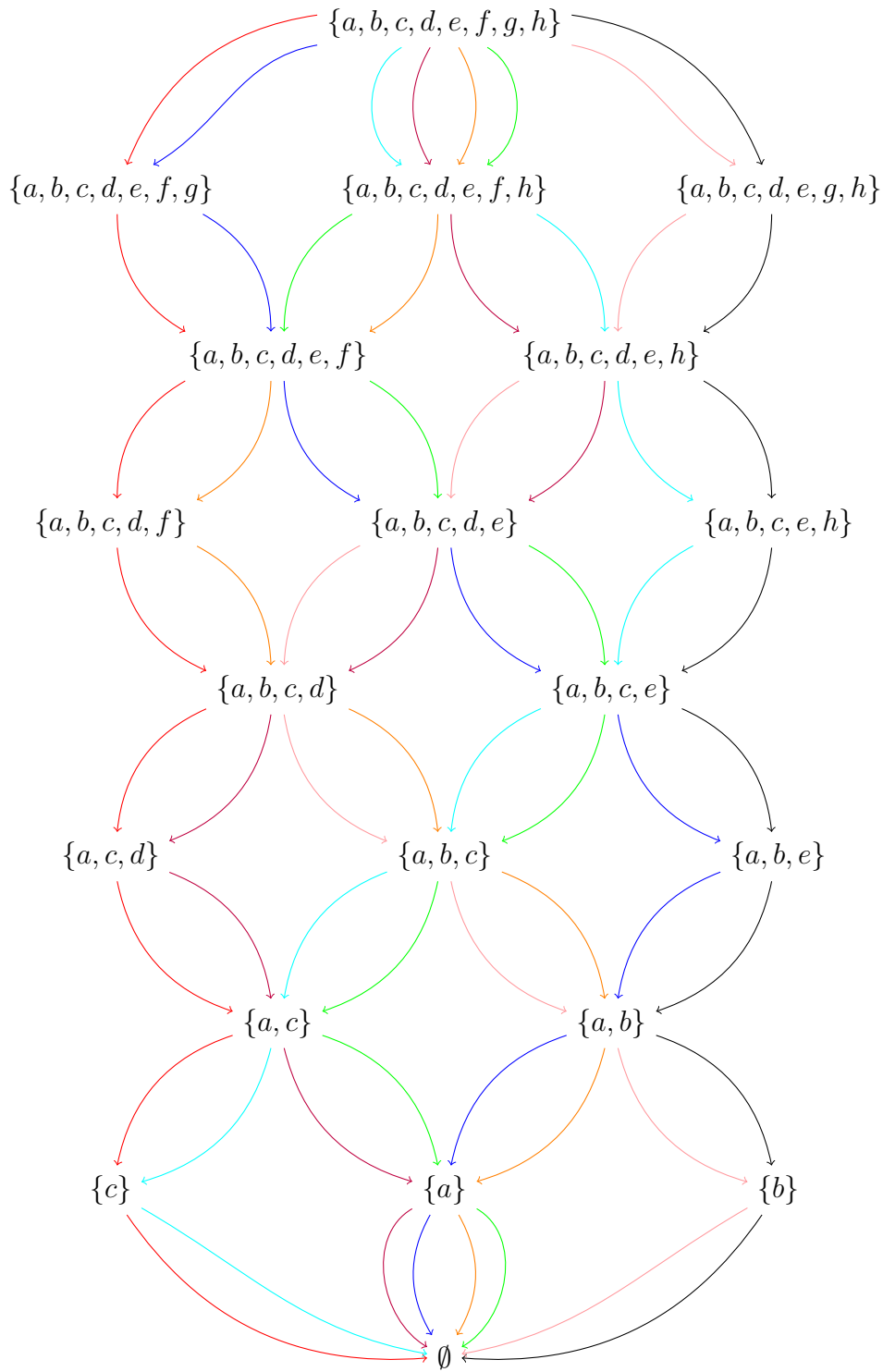
\begin{figure}
    \centering
    \begin{tikzpicture}[scale=.4,->]
    \tikzstyle{every node} = [rectangle]
        \node (a) at (0,0) {$\emptyset$};

        \node (b) at (-12,6) {$\{c\}$};
        \node (c) at (0,6) {$\{a\}$};
        \node (d) at (12,6) {$\{b\}$};

        \node(e) at (-6,12) {$\{a,c\}$};
        \node(f) at (6,12) {$\{a,b\}$};
        
        \node (g) at (-12,18) {$\{a,c,d\}$};
        \node (h) at (0,18) {$\{a,b,c\}$};
        \node (i) at (12,18) {$\{a,b,e\}$};
        
        \node (j) at (-6,24) {$\{a,b,c,d\}$};
        \node (k) at (6,24) {$\{a,b,c,e\}$};

        \node (l) at (-12,30) {$\{a,b,c,d,f\}$};
        \node (m) at (0,30) {$\{a,b,c,d,e\}$};
        \node (n) at (12,30) {$\{a,b,c,e,h\}$};

        \node (o) at (-6,36) {$\{a,b,c,d,e,f\}$};
        \node (p) at (6,36) {$\{a,b,c,d,e,h\}$};

        \node (q) at (-12,42) {$\{a,b,c,d,e,f,g\}$};
        \node (r) at (0,42) {$\{a,b,c,d,e,f,h\}$};
        \node (s) at (12,42) {$\{a,b,c,d,e,g,h\}$};

        \node (t) at (0,48) {$\{a,b,c,d,e,f,g,h\}$};
        

        \path (t) edge [bend left] (s);
        \path (s) edge [bend left] (p);
        \path (p) edge [bend left] (n);
        \path (n) edge [bend left] (k);
        \path (k) edge [bend left] (i);
        \path (i) edge [bend left] (f);
        \path (f) edge [bend left] (d);
        \path (d) edge [bend left] (a);

        \path (t) edge [bend right, red] (q);
        \path (q) edge [bend right, red] (o);
        \path (o) edge [bend right, red] (l);
        \path (l) edge [bend right, red] (j);
        \path (j) edge [bend right, red] (g);
        \path (g) edge [bend right, red] (e);
        \path (e) edge [bend right, red] (b);
        \path (b) edge [bend right, red] (a);

        \path (t) edge [out=350, in=150, amaranth] (s);
        \path (s) edge [bend right, amaranth] (p);
        \path (p) edge [bend right, amaranth] (m);
        \path (m) edge [bend right, amaranth] (j);
        \path (j) edge [bend right, amaranth] (h);
        \path (h) edge [bend right, amaranth] (f);
        \path (f) edge [bend right, amaranth] (d);
        \path (d) edge [out=210, in=10, amaranth] (a);

        \path (t) edge [out=190, in=30, blue] (q);
        \path (q) edge [bend left, blue] (o);
        \path (o) edge [bend right, blue] (m);
        \path (m) edge [bend right, blue] (k);
        \path (k) edge [bend right, blue] (i);
        \path (i) edge [bend right, blue] (f);
        \path (f) edge [bend right, blue] (c);
        \path (c) edge [bend right, blue] (a);

        \path (t) edge [bend left, orange] (r);
        \path (r) edge [bend left, orange] (o);
        \path (o) edge [bend left, orange] (l);
        \path (l) edge [bend left, orange] (j);
        \path (j) edge [bend left, orange] (h);
        \path (h) edge [bend left, orange] (f);
        \path (f) edge [bend left, orange] (c);
        \path (c) edge [bend left, orange] (a);

        \path (t) edge [out=330, in=30, green] (r);
        \path (r) edge [bend right, green] (o);
        \path (o) edge [bend left, green] (m);
        \path (m) edge [bend left, green] (k);
        \path (k) edge [bend left, green] (h);
        \path (h) edge [bend left, green] (e);
        \path (e) edge [bend left, green] (c);
        \path (c) edge [out=330, in=30, green] (a);

        \path (t) edge [bend right, purple] (r);
        \path (r) edge [bend right, purple] (p);
        \path (p) edge [bend left, purple] (m);
        \path (m) edge [bend left, purple] (j);
        \path (j) edge [bend left, purple] (g);
        \path (g) edge [bend left, purple] (e);
        \path (e) edge [bend right, purple] (c);
        \path (c) edge [out=210, in=150, purple] (a);

        \path (t) edge [out=210, in=150, aqua] (r);
        \path (r) edge [bend left, aqua] (p);
        \path (p) edge [bend right, aqua] (n);
        \path (n) edge [bend right, aqua] (k);
        \path (k) edge [bend right, aqua] (h);
        \path (h) edge [bend right, aqua] (e);
        \path (e) edge [bend left, aqua] (b);
        \path (b) edge [out=330, in=170, aqua] (a);

    \end{tikzpicture}
    \caption{The graphical representation of Example \ref{counterexample}. Each differently colored path corresponds to a different preference in our counterexample.}
    \label{fig:Counterexample}
\end{figure}

We now interpret our definition of edge decomposability in terms of the probability flow diagram. Recall that every pair $(x,A)$ with $x \in A$ is associated with a specific edge of the probability flow diagram. Specifically, $q(x,A)$ is assigned as the edge capacity of the edge associated with $(x,A)$. Edge decomposability then says, given a model $M$, for each submodel $N$ there exists some path associated with $\succ \in N$ such that this path has some edge which is unique to it among paths associated with preferences $\succ' \in N$. Following a similar process to what was described in the last paragraph, this means that when a graph is written as a combination of flows along paths associated with preferences in $\succ$ in a edge decomposable model $M$, we can always uniquely decompose this graph into flow weights on paths associated with preferences in $M$.\footnote{Further, we can think of the analogue of edge decomposability for graphs other than the probability flow diagram. In this case, edge decomposability still guarantees that we are able to uniquely decompose our graph into a series of flows into our edge decomposable set of paths.} While edge decomposability may be sufficient for identification of a model $M$, it fails to be necessary. We show this through Example \ref{counterexample} and Figure \ref{fig:Counterexample}.

\begin{ex}\label{counterexample}
    Consider the environment given by $X_n=\{a,b,c,d,e,f,g,h\}$. We now provide a model with eight preferences which is not edge decomposable and yet is identified. These preferences are represented on the probability flow diagram of Figure \ref{fig:Counterexample}. The paths and preferences are as follows.
    \begin{enumerate}
        \item {\color{black}Black} path - $\succ_1$ - $f \succ g \succ d \succ h \succ c \succ e \succ a \succ b$
        \item {\color{red}Red} path - $\succ_2$ - $h \succ g \succ e \succ f \succ b \succ d \succ a \succ c$
        \item {\color{amaranth}Pink} path - $\succ_3$ - $f \succ g \succ h \succ e \succ d \succ c \succ b \succ a$
        \item {\color{blue}Blue} path - $\succ_4$ - $h \succ g \succ f \succ d \succ c \succ e \succ b \succ a$
        \item {\color{aqua}Aqua} path - $\succ_5$ - $g \succ f \succ d \succ h \succ e \succ b \succ a \succ c$
        \item {\color{green}Green} path - $\succ_6$ - $g \succ h \succ f \succ d \succ e \succ b \succ c \succ a$
        \item {\color{purple}Purple} path - $\succ_7$ - $g \succ f \succ h \succ e \succ b \succ d \succ c \succ a$
        \item {\color{orange}Orange} path - $\succ_8$ - $g \succ h \succ e \succ f \succ d \succ c \succ b \succ a$
    \end{enumerate}
    We now proceed to show that $M=\{\succ_1,\succ_2,\succ_3,\succ_4,\succ_5,\succ_6,\succ_7,\succ_8\}$ is an identified model despite not being edge decomposable. It follows immediately from Figure \ref{fig:Counterexample} that $M$ is not edge decomposable as each edge covered by $M$ is covered by two paths. Now we show identification by directly calculating the values of $\nu$. To begin, we start by identifying the probability weight on the preferences $\succ_2$, $\succ_4$, and $\succ_8$.
    \begin{equation}
        \begin{split}
            \nu(\succ_2)+\nu(\succ_4) & = q(h,\{a,b,c,d,e,f,g,h\}) \\
            \nu(\succ_2)+\nu(\succ_8) & = q(e,\{a,b,c,d,e,f\}) \\
            \nu(\succ_4)+\nu(\succ_8) & = q(b,\{a,b\})
        \end{split}
    \end{equation}
    This gives us the following.
    \begin{equation}\label{firstident}
        \begin{split}
            \nu(\succ_2) & = \frac{q(h,\{a,b,c,d,e,f,g,h\})+q(e,\{a,b,c,d,e,f\})-q(b,\{a,b\})}{2} \\
            \nu(\succ_4) & = \frac{q(h,\{a,b,c,d,e,f,g,h\})-q(e,\{a,b,c,d,e,f\})+q(b,\{a,b\})}{2} \\
            \nu(\succ_8) & = \frac{-q(h,\{a,b,c,d,e,f,g,h\})+q(e,\{a,b,c,d,e,f\})+q(b,\{a,b\})}{2}
        \end{split}
    \end{equation}
    The probability weight on each other preference is then given by the following.
    \begin{equation}\label{lastident}
        \begin{split}
            \nu(\succ_1) & = q(c,\{a,b,c,e\})-\nu(\succ_4) \\
            \nu(\succ_6) & = q(f,\{a,b,c,d,e,f\})-\nu(\succ_4) \\
            \nu(\succ_3) & = q(a,\{a,b\})-\nu(\succ_1) \\
            \nu(\succ_5) & = q(e,\{a,b,c,e\})-\nu(\succ_6) \\
            \nu(\succ_7) & = q(c,\{a,c\})-\nu(\succ_6) \\
        \end{split}
    \end{equation}
    Note that each line of Equation \ref{lastident} is written only using variables that have been identified from Equation \ref{firstident} or higher lines in Equation \ref{lastident}. Equations \ref{firstident} and \ref{lastident} tell us we can recover our distribution over preferences $\nu$ directly from the data for model $M$, and so $M$ is identified despite not being edge decomposable.
\end{ex}

As Example \ref{counterexample} demonstrates, edge decomposability is not necessary for identification. However, edge decomposability is a condition that is relatively easy to use. By that we mean that it is easy to constructively create edge decomposable models. Simply start with a preference $\succ$ and consider the model $N=\{\succ\}$. Given $N$, find a pair $(x,A)$ with $x \in A$ such that $N \cap L(x,A)=\emptyset$. Choose any $\succ' \in L(x,A)$ and create a larger model $N \cup \{\succ'\}$. We can repeat this process until no such pairs $(x,A)$ exist. Once no such pairs exist, we have a maximal edge decomposable model. In fact, as Theorem \ref{edgeIdentification} states, maximal edge decomposable models are as large as maximal identified models. However, in order to show this, we need to first introduce two algorithms.

In graph theory, a common question is finding a cycle basis for a given graph. The standard construction is to find a spanning tree for that graph and then construct a single cycle for each edge in the graph which is not in the spanning tree. A spanning tree of a graph is a tree that visits every node of that graph. A tree is a connected undirected graph with no (undirected) cycles. The standard construction develops a cycle basis, but these cycles are not guaranteed to be circuits. Recall that preferences correspond to minimal circuits in the appended probability flow diagram. This means that we are interested in constructing a basis for the circuit space using minimal circuits. As such, we are unable to apply the standard construction right out of the box.

Our first algorithm takes the appended probability flow diagram and constructs a spanning tree which respects the direction of each edge. The fact that each edge in the spanning tree respects the direction of the edges in our original graph is important for our second algorithm. Our second algorithm takes in the spanning tree and creates a cycle basis using minimal circuits. In other words, our second algorithm uses the output of our first algorithm to create our preference basis. The fact that we are able to use minimal circuits in our second algorithm is a result of the specific construction we use in our first algorithm. We now introduce our first algorithm and show that it constructs a spanning tree which respects the direction of edges in the appended probability flow diagram.

\begin{algo}\label{treealgo}
    \begin{enumerate}
        \item The root node of our spanning tree is chosen as $\emptyset$.
        \item Connect $\emptyset$ with $X_n$ via a directed edge. Set $i= n$.
        \item For the sets $A$ with $|A|=i$, enumerate these sets from $1$ to $J$. Set $j=1$.
        \item For each directed (out-)edge, $e(A_j,B)$ leaving $A_j$, add (the directed edge) $e(A_j,B)$ to our tree if $B$ is not already connected to our tree.
        \item Set $j=j+1$. If $j \leq J$, return to the previous step. If $j > J$, proceed to the next step.
        \item Set $i=i-1$. If $i\leq 0$, terminate the algorithm. If $i\geq 1$, return to step 3 of the algorithm.
    \end{enumerate}
\end{algo}

\begin{prop}
    Algorithm \ref{treealgo} creates a spanning tree of the probability flow diagram. Further, the edges of the spanning tree respect the direction of the edges in the probability flow diagram.
\end{prop}

\begin{proof}
    We first prove that this algorithm constructs a spanning tree. Suppose towards a contradiction that we do not. There are three cases.
    \begin{enumerate}
        \item The graph is not spanning. \\
        This means that there is some node $A$ in the probability flow diagram which is not in our tree. This means that every parent node of $A$ in the probability flow diagram is also not in our tree (by step 4 of the algorithm).\footnote{A parent node of $A$ is any node such that there exists a directed edge from the node to $A$.} Recursively applying this argument eventually takes us to node $X_n$ which is in our tree by construction. This is a contradiction.
        \item The graph is not connected. \\
        By the previous case, we know that every node $A$ is connected to some parent node. Repeated application of this logic implies that every node $A$ is connected to $X_n$ via a path which is also connected to $\emptyset$. Thus the graph is connected.
        \item There exists some cycle in the constructed graph. \\
        Since our graph is connected, this means that there are two distinct paths between $X_n$ and some node $A$. By step 4 of our algorithm, the indegree of $A$ is equal to one.\footnote{Note that a tree corresponds to the undirected version of the constructed graph. We however included directed edges in our construction as they will be useful in some arguments going forward.} This means that there is a single parent node of $A$. Call this parent node $B$. Repeat this argument applied to $B$ and then recursively until we reach $X$. This means that there is a single path going from $A$ to $X_n$ which goes from child node to parent node.\footnote{A node $A$ is a child node of $B$ if $B$ is a parent node of $A$.} Thus if there are two paths from $A$ to $X_n$, without loss, the second path at some point goes from parent node to child node. However, for every child node, there is also a unique path going from that node to $X_n$ that goes from child to parent nodes. Repeating this argument, we eventually get to some node of the form $\{x\}$. There are no edges on our tree directly connecting $\{x\}$ to $\emptyset$ (by steps 1,2 and 4 in the algorithm), thus there is a unique path from $\{x\}$ to $X_n$ and we are done.
    \end{enumerate}
    As none of our three cases can hold, we have a contradiction. Thus our algorithm creates a spanning tree. Further, by the fact that the algorithm initializes going from $\emptyset$ to $X_n$ and then, at each step of the algorithm, starts at $A$ and goes to $A \setminus \{x\}$ for some $x \in A$, the constructed spanning tree respects the direction of edges in the appended probability flow diagram.
\end{proof}

We now introduce our second algorithm which takes in a spanning tree created from Algorithm \ref{treealgo} and creates a preference basis.

\begin{algo}\label{prefbasis}
\begin{enumerate}
    \item Take as input the spanning (directed) tree from Algorithm \ref{treealgo} and the appended probability flow diagram. Initialize with $i=1$.
    \item Enumerate the sets $A$ with $|A|=i$ from $1$ to $J$. Set $j=1$.
    \item Enumerate from $1$ to $K$ each directed edge leaving $A_j$ in the probability flow diagram which is not a part of our appended spanning tree. Set $k=1$. If $K=0$, proceed to step 6.
    \item Add directed edge $e_k$ to the appended spanning tree. By definition of spanning tree, the addition of this edge creates a (new) cycle. \begin{cl}\label{algoclaim}
        This cycle can be chosen to be a circuit.
    \end{cl}
    \item Since this cycle can be chosen to be a circuit, it can be chosen to be a minimal circuit.\footnote{This is because a minimal circuit corresponds to a path through the appended probability flow diagram plus the edge connecting $\emptyset$ with $X$.} Minimal circuits in the probability flow diagram correspond to preferences. Add the corresponding preference to the set $\{\succ_l\}_{l=1}^L$
    \item Set $k=k+1$. If $k\leq K$, return to step 4. If not, proceed to the next step.
    \item Set $j=j+1$. If $j \leq J$, return to step 3. If not proceed to the next step.
    \item Set $i=i+1$. If $i \leq |X|$, return to step 2. If not, terminate the algorithm.
\end{enumerate}
    
\end{algo}

We now proceed with a proof of Claim \ref{algoclaim}.

\begin{proof}
    We proceed by induction on $i$. Suppose that $i=1$. In this case, our edge is connecting $\{x\}$ to $\emptyset$. By construction of our spanning tree in Algorithm \ref{treealgo} (step 4), there is a path from $\{x\}$ to $X_n$ in the un-appended spanning tree that goes from children nodes to parent nodes in each step. Since $\emptyset$ is directly connected $X_n$, by adding the edge connecting $\{x\}$ to $\emptyset$, we create a circuit.

    Now suppose that every edge leaving nodes $A$ where $|A| < i$ has been added to the appended spanning tree. We want to show that adding an edge leaving node $B$ with $|B|=i$ forms some circuit. By a similar argument, there is a directed path from $X_n$ to $B$ in our appended spanning tree. This path goes from parent to children nodes and respects the direction of edges from the probability flow diagram. An edge leaving $B$ connects to a node $A$ with $|A|=i-1$. Since every edge leaving nodes of size $i-1$ or less have been added to our appended spanning tree, there exists a path from $A$ to $\emptyset$ which respects the direction of edges from the probability flow diagram. By adding the edge connecting $B$ to $A$, there is a directed path from $X_n$ to $B$, from $B$ to $A$, from $A$ to $\emptyset$, and from $\emptyset$ to $X_n$. This is a minimal circuit, and so we are done.
\end{proof}

\begin{prop}\label{edgeSizeProp}
    The set of preferences $\{\succ_l\}_{l=1}^{|M|}$ created by Algorithms \ref{treealgo} and \ref{prefbasis} are (sequentially) edge decomposable. Further, $|M|=(n-2)2^{n-1}+2$.
\end{prop}

\begin{proof}
    Observe that in each iteration of Algorithm \ref{prefbasis}, we create a (minimal) circuit using only edges from the spanning tree, the edge we just added, and edges that were added in prior iterations of the algorithm. This means that the edge we are adding in a specific step of an algorithm is being used for the first time in that step of the algorithm. Each of these edges corresponds to a tuple $(x,A)$. Let $(x_k,A_k)$ correspond to the edge from the $k$th iteration of the algorithm. By the prior argument, this means that $\{\succ_l\}_{l=1}^K \cap L(x_K,A_k)=\{\succ_K\}$ for $K \leq |M|$. This is exactly the definition of sequential edge decomposability.

    Now we show that $|M|=(n-2)2^{n-1}+2$. $|M|$ corresponds to the number of edges in the probability flow diagram which \textit{are not} in the spanning tree. A standard result in graph theory (see \citet{berge2001theory} p. 29 Theorem 2) is that the cyclomatic number of a graph is equal to the size of its cycle basis. Further, the standard construction of cycle bases is via constructing a spanning tree and then using each edge \textit{not} in the spanning tree to create a set of cycles (see again \citet{berge2001theory} or more explicitly \citet{bollobas1998modern} p. 53 Theorem 9). We have done this in our Algorithms \ref{treealgo} and \ref{prefbasis}. We have constructed a set of circuits with one circuit for each edge not in our spanning tree. Thus the number of circuits, and thus preferences in the set $\{\succ_l\}_{l=1}^{|M|}$, we have created is equal to the cyclomatic number of the probability flow diagram. But this is exactly $(n-2)2^{n-1}+2$ by our main theorem, and so we are done.
\end{proof}

It then follows from the paragraph preceding Theorem \ref{edgeIdentification}, Example \ref{counterexample}, and Proposition \ref{edgeSizeProp} that Theorem \ref{edgeIdentification} holds. In effect, Theorem \ref{edgeIdentification} tells us that edge decomposability still allows us to form maximal identified models. It simply puts a restriction on which of these models are allowed.

\subsection{Relationship to Other Uniqueness Conditions}
We now compare edge decomposability to three other uniqueness conditions. Two of these conditions come from the random utility literature and the third comes from Example \ref{latinsquare1}. The two conditions from the literature are the uniqueness conditions introduced in \citet{turansick2022identification} and \citet{apesteguia2017single}. While both of these restrictions recover uniqueness in different ways, they both work by placing restrictions on the preferences in the induced support of their model. Our main goal is to compare these restrictions on supports to our edge decomposability condition on models. We will show that the supports induced by \citet{turansick2022identification}, \citet{apesteguia2017single}, and the model from Example \ref{latinsquare1} are subsets of the possible supports/models induced by edge decomposability.

To begin, we review the setup and characterization of \citet{turansick2022identification}. The goal of \citet{turansick2022identification} is to characterize when the full model $M=\mathcal{L}(X_n)$ admits a unique representation. Unlike our focus, which is developing ex ante conditions that imply identification, the focus of \citet{turansick2022identification} is to consider the unrestricted random utility model and characterize which data sets have a unique rationalizing distribution. In the proof of Theorem 1 from \citet{turansick2022identification}, it is shown that a random choice rule $p$ has a unique representation if and only if every supported path in the probability flow diagram has some edge unique to that path among supported paths. Here, by supported paths, we mean paths where every edge on that path has a strictly positive edge capacity. Restating this in term of preferences, this means that $p$ has a unique support if and only if one of the following conditions hold for each preference.
\begin{enumerate}
    \item $\succ \in L(x,A)$ and $q(x,A)=0$
    \item $\exists (x,A)$ with $x\in A$ such that $L(x,A)\cap\{\succ'|\succ' \in L(y,B) \implies q(y,B) > 0\}=\{\succ\}$
\end{enumerate}
Relating this back to our definition of edge decomposability, let $M=\{\succ'|\succ' \in L(y,B) \implies q(y,B) > 0\}$. Then the result of \citet{turansick2022identification} is asking that for every $\succ \in M$, there exists some $(x,A)$ with $x \in A$ such that $M \cap L(x,A) = \{\succ\}$. The condition of \citet{turansick2022identification} implies edge decomposability. We now show through an example, related to the counterexample of \citet{fishburn1998stochastic}, that it is strictly stronger than edge decomposability.

\begin{ex}\label{singlefishburn}
    Consider the model $M$ given by the following three preferences.
    \begin{enumerate}
        \item $a \succ b \succ c \succ d=\succ_1$
        \item $b \succ a \succ d \succ c=\succ_2$
        \item $a \succ b \succ d \succ c=\succ_3$
    \end{enumerate}
    Observe the following.
    \begin{enumerate}
        \item $M \cap L(a,\{a,b,c,d\})=\{\succ_1,\succ_3\}$
        \item $M \cap L(b,\{b,c,d\})=\{\succ_1,\succ_3\}$
        \item $M \cap L(c,\{c,d\})=\{\succ_2,\succ_3\}$
        \item $M \cap L(d,\{d\})=\{\succ_2,\succ_3\}$
    \end{enumerate}
    This means that $\succ_3$ never has a set $L(x,A)$ with $\succ_3 \in L(x,A)$ that is unique to $\succ_3$ among preferences in $M$. However, as $M\cap L(b,\{a,b,c,d\})=\succ_2$ and $M \cap L(d,\{c,d\})=\{\succ_1\}$, it is easy to see that $M$ is edge decomposable.
\end{ex}

We now consider the single crossing random utility model (SCRUM) of \citet{apesteguia2017single}. SCRUM puts further structure on $X_n$ in that SCRUM assumes that $X_n$ is endowed with some exogenous linear order $\rhd$. We say that a random choice rule $p$ is rationalizable by SCRUM if there exists a distribution over preferences $\nu$ such that the support of $\nu$ can be ordered so that it satisfies the single-crossing property with respect to $\rhd$. Recall from Example \ref{scrum1} the single-crossing property.
\begin{dfn}
We say that a distribution over preferences $\nu$ satisfies the single-crossing property if the support of $\nu$ can be ordered in such a way that for all $x \rhd y$, $x \succ_i y$ implies $x \succ_j y$ for all $j \geq i$.
\end{dfn}
Unlike previously, it is not immediately obvious that the supports of SCRUM representations are edge decomposable. In order to see that they are, note the following.
\begin{prop}\label{SCRUMEdge}
    Suppose that $\nu$ is a distribution over preferences whose support is ordered $(\succ_1,\dots,\succ_n)$ so that it satisfies the single-crossing property with respect to $\rhd$. Then there exists a pair $(x,A)$ with $x \in A$ such that $L(x,A) \cap \{\succ_1,\dots,\succ_n\}=\{\succ_1\}$.
\end{prop}

It follows from Proposition \ref{SCRUMEdge} that the supports of SCRUM representations satisfy edge decomposability. Specifically, we know that every subset of a SCRUM support is itself the support of some SCRUM representation. It then follows that the lowest ranked preference in the support is the unique element of some $L(x,A)$ among preferences in the support. In other words, once the support of a SCRUM representation is pinned down, we can recursively find the probability weights on the preferences by looking at the lowest and then the next lowest ranked preferences in the support. We now show through an example that SCRUM supports fail to capture every edge decomposable model.

\begin{ex}\label{DoubleFisburn}
    Let $X_n=\{a,b,c,d,e,f,g,h\}$ be endowed with the linear order $\rhd$ and suppose that $M$ is given by the following.
        \begin{enumerate}
        \item $a \succ b \succ c \succ d \succ f \succ e = \succ_1$
        \item $a \succ b \succ d \succ c \succ e \succ f = \succ_2$
        \item $b \succ a \succ c \succ d \succ e \succ f = \succ_3$
    \end{enumerate}
    We now proceed with some case work.
    \begin{enumerate}
        \item Suppose $a \rhd b$. Then it is the case that, abusing notation, if $M$ can satisfy the single-crossing property, $\succ_1 \rhd \succ_3$ and $\succ_2 \rhd \succ_3$.
        \begin{enumerate}
            \item Suppose $c \rhd d$. Then it is the case that $\succ_1 \rhd \succ_2$ and $\succ_3 \rhd \succ_2$. However, given that $a \rhd b$ and $\succ_2 \rhd \succ_3$, $c \rhd d$ cannot be the case.
            \item Suppose $d \rhd c$. This means that $\succ_2 \rhd \succ_1$ which further means that $e \rhd f$. However, as $a \rhd b$ and $\succ_1 \rhd \succ_3$, we also have that $f \rhd e$, and so $d \rhd c$ cannot be the case.
        \end{enumerate}
        \item Suppose $b \rhd a$. This means that $\succ_3 \rhd \succ_1$ and $\succ_3 \rhd \succ_2$. This gives us that $a\rhd b$, $c \rhd d$, and $e \rhd f$. As $c \succ_1 d$, $d \succ_2 c$, $f \succ_1 e$, and $e \succ_2 f$, we have that $\succ_1 \rhd \succ_2 \rhd \succ_1$, which is a cycle. Thus $b \rhd a$ cannot be the case.
    \end{enumerate}
    The above case work tells us that it can neither be the case that $a \rhd b$ nor $b \rhd a$, and thus $\{\succ_1,\succ_2,\succ_3\}$ cannot be a SCRUM support. However, as $\succ_1$ is the only preference to rank $f \succ e$, $\succ_2$ is the only preference to rank $d \succ c$, and $\succ_3$ is the only preference to rank $b \succ a$, the model $M=\{\succ_1,\succ_2,\succ_3\}$ is edge decomposable.
\end{ex}

We note that while edge decomposability is a strictly weaker condition on supports than single-crossing, part of the novelty of SCRUM is that there is an endogenous mapping from $\rhd$ and $p$ to the support of $\nu$. Alternatively, we could have relied on Theorem \ref{edgeIdentification} and Proposition \ref{scrumSize} to show that edge decomposability is a strictly weaker condition than the single-crossing condition on models. Overall, our discussion in this section shows that edge decomposability is a weaker identifying restriction on supports than two recent identifying restrictions. Further, edge decomposability is an easy to check and constructive criterion for identification.

\subsubsection{Preference Aggregation and Latin Squares}\label{sec:CARUM}
In this section we formalize and discuss the model described in Example \ref{latinsquare1}. Importantly, we show that this model ends up being edge decomposable and thus identified. Let $\rhd$ be a linear order on $X_n$. Without loss we can have $1 \rhd 2 \rhd \dots \rhd n$.

\begin{dfn}
    We say that a preference $\succ$ \textbf{respects} $\rhd$ if there exists $m\in \{1,\dots,n\}$
    such that $m \succ m+1 \succ \dots \succ n-1 \succ n \succ 1 \succ \dots \succ m-1$.
\end{dfn}

The respects relationship is an equivalence relation, whereby any two linear orders inducing the same Latin square are equivalent. Thus, given an order $\rhd$, we can think of a full support distribution on preferences which respect $\rhd$ as the classic example in social choice theory that leads to Condorcet cycles when aggregating preferences.\footnote{We note that there has been recent interest in connecting social choice theory, preference aggregation, and stochastic choice. \citet{brandl2016consistent} studies aggregation of distributions over preferences and characterizes which aggregation rules satisfy a consistency condition when combining populations and a consistency condition when varying the available menu. \citet{sprumont2024randomized} studies aggregation via randomized tournaments and gives conditions under which there is a unique extension to a full random choice rule.} To extend this further, given an ordering $\rhd=1 \rhd 2 \rhd \dots \rhd n$, we can think of appending this ordering to make the cycle $1 \rhd 2 \rhd \dots \rhd n \rhd 1$. This cycle corresponds to the Condorcet cycle induced when aggregating preferences according to an $n-1$ majority rule, which respects $\rhd$.

\begin{dfn}
We say that a model $M$ is a Condorcet Aggregate Random Utility Model (CARUM) if the following conditions hold for some $\rhd$:
    \begin{itemize}
        \item Each $\succ\in M$ respects $\rhd$.
        \item Each $\succ$ which respects $\rhd$ satisfies $\succ\in M$.
    \end{itemize}
    For a given $\rhd$, denote the corresponding CARUM by $M(\rhd)$.
\end{dfn}

The set of CARUM models enjoys a particularly interesting identification property.  Assume that the analyst has knowledge that a given random choice rule $p$ is consistent with CARUM, but lacks knowledge about the ordering $\rhd$ of alternatives.  It turns out that, up to orders that induce the same Latin square, the actual $\rhd$ can be identified.  This is the content of the following result.

\begin{prop}\label{carumIdent}For each $\rhd$, $M(\rhd)$ is edge decomposable.  Further, for any $\rhd$ and $\rhd'$, any random choice rule $p$, and any $\nu\in\Delta(M(\rhd))$ and $\nu'\in\Delta(M(\rhd'))$ for which $p=p_{\nu}=p_{\nu'}$, we have $M(\rhd)=M(\rhd')$ and $\nu=\nu'$.\end{prop}

The first part of this Proposition \ref{carumIdent} is easy to see. In a given Latin square $M(\rhd)$, there is exactly one preference $\succ$ which ranks $x$ first for each $x \in X_n$. Thus $L(x,X_n)\cap M(\rhd)=\{\succ\}$ for each $x$ and we have edge decomposability. The intuition of the second part of Proposition \ref{carumIdent} is as follows. Fix a given Latin square $M(\rhd)$. For each $A \neq X_n$, there is at most a single alternative $x \in A$ such that $q(x,A) >0$. Consider some edge connecting $X_n$ to $X_n\setminus \{x\}$ in the probability flow diagram. If $q(x,X_n)>0$, the prior observation implies that there is a unique path from $X_n\setminus \{x\}$ to $\emptyset$ with strictly positive edge capacities along each edge of the path. This allows us to say that the preference $\succ$ which corresponds to this path is in the support of any rationalizing distribution. Since respects is an equivalence relation, we can choose $\rhd$ to be $\succ$. From here we simply construct $M(\rhd)$ and apply edge decomposability to recover identification of $\nu$.

\section{Related Literature}\label{RelLit}

Our paper is primarily related to two strands of literature. The first strand is the one which studies uniqueness and identification in random utility models. Study of the random utility model goes back to \citet{block1959random} and \citet{falmagne1978representation}. \citet{barbera1986falmagne} and \citet{fishburn1998stochastic} note that the random utility model is in general not identified when there are at least four alternatives available. Much of the literature studying identification aims to recover uniqueness by refining the random utility model. The random expected utility model of \citet{gul2006random} recovers uniqueness by restricting to expected utility functions while studying choice over lotteries. As discussed earlier, the single-crossing random utility model of \citet{apesteguia2017single} is able to recover uniqueness by asking that the support of their representation satisfy the single-crossing property with respect to some exogenous order. \citet{yildiz2023foundations} extends this line of thought and characterizes exactly which models of choice can be identified through a similar progressivity condition. Another approach to identification is through the use of stronger data. \citet{lu2019bayesian} is able to recover both beliefs and preferences when stochastic choice data and information sources are observed. \citet{dardanoni2020inferring} is able to recover preferences as well as cognitive heterogeneity with data that connects agents' choices across menus. An alternative approach is taken by \citet{turansick2022identification}. \citet{turansick2022identification} simply asks which realizations of standard random choice rules admit a unique random utility representation. \citet{azrieli2022marginal} studies stochastic choice and random utility when we are unable to condition choice on the menu of available alternatives. They find that in this setting that the random utility model is also unidentified. \citet{kashaev2024entangled} studies separable stochastic choice and finds that separability can be characterized by each agent having well defined marginal choices if and only if the set of feasible choice functions (i.e. preferences) are linearly independent.

The second strand of literature our paper is related to is the strand which uses graph theoretic tools to study the random utility model and more generally stochastic choice. To our knowledge, this strand of literature was started by \citet{fiorini2004short} who used the observation that preferences can be represented as flows on the probability flow diagram in order to provide an alternative proof of the characterization of \citet{falmagne1978representation}. \citet{turansick2022identification} studies conditions on the probability flow diagram which characterize when random choice rules have a unique random utility representation. \citet{chang2022approximating} studies which preferences have adjacent paths in the probability flow diagram in order to say when random-coefficient models are good approximations of the random utility model. \citet{chambers2024correlated} extends the probability flow diagram to multiple dimension in order to study choice across multiple people or time periods. Finally, \citet{kono2023axiomatization} use graph theoretic tools to study the random utility model when choice probabilities of some alternatives are unobserved.

\appendix

\section{Omitted Proofs}

\subsection{Proof of Lemma \ref{Plinind}}
\begin{proof} If $\{p_\succ\}_{\succ\in M}$ is not linearly independent, then for each $\succ\in M$, there is $c_{\succ}\in\mathbb{R}$ for which $\sum_{\succ\in M}c_{\succ}p_{\succ}=0$, where at least one $c_{\succ}\neq 0$.  As each $p_{\succ}\geq 0$ componentwise, it follows that $\{\succ\in M:c_{\succ}>0\}\neq \varnothing$ and $\{\succ\in M:c_{\succ}<0\}\neq \varnothing$.  Observe then that $\sum_{\{\succ\in M:c_{\succ}>0\}}c_{\succ}p_{\succ} = \sum_{\{\succ\in M:c_{\succ}<0\}}-c_{\succ}p_{\succ}$.  Now, define $\nu\in \Delta(M)$ as $\nu(\succ')=\frac{c_{\succ'}}{\sum_{\{\succ\in M:c_{\succ}>0\}}c_{\succ}}$ when $c_{\succ'}>0$, and $\nu(\succ')=0$ otherwise.  Similarly define $\nu'\in \Delta(M)$ as $\nu(\succ')=\frac{-c_{\succ'}}{\sum_{\{\succ\in M:c_{\succ}>0\}}-c_{\succ}}$ when $c_{\succ'}<0$ and $\nu(\succ')=0$ otherwise.  Observe that $\nu$ and $\nu'$ have disjoint supports, but by construction $p_{\nu}=p_{\nu'}$, so that $M$ is not identified.  If $\{p_\succ\}_{\succ\in M}$ is linearly independent, it is easy to see that $M$ is identified:  $p_{\nu}=p_{\nu'}$ implies $\sum_{\succ\in M}(\nu(\succ)-\nu'(\succ))p_{\succ}=0$, so that $\nu(\succ)-\nu'(\succ)=0$ for all $\succ\in M$ by the definition of linear independence.\end{proof}

\subsection{Proof of Lemma \ref{qLinIndlem}}

\begin{proof}
    \citet{rota1964foundations}, Propositions 1 and 2, shows that there is a bijective and linear relationship between a M\"{o}bius inverse $q$ and its generating function $p$. This bijection is through Equation \ref{mobinvform}. Hence, there is an invertible linear map carrying generating functions $p$ to M\"{o}bius inverses $q$.  The result follows as linear independence is preserved under invertible linear transformations.
\end{proof}

\subsection{Proof of Proposition \ref{add1prop}}

\begin{proof}
One direction is obvious (the linear independence of $\{x^i\}_{i=1}^I$ entails the linear independence of $\{(x^i,1)\}_{i=1}^I$).  For the other direction, suppose by means of contradiction that $\{(x^i,1)\}_{i=1}^K$ is linearly independent but that there are $c^i$ not all $0$ for which $\sum_i c^i x^i = 0$.  Let us suppose without loss that $\sum_k x^i_k = 1$ for all $i$.  Observe then that $\sum_k\sum_i c^i x^i_k=0$, and that $\sum_k \sum_i c^i x^i_k = \sum_ic^i \sum_k x^i_k =\sum_i c^i$.  Consequently $\sum_i c^i  = 0$ and hence $\sum_i c^i (x^i,1) = 0$, contradicting linear independence of $\{(x^i,1)\}_{i=1}^I$. 
\end{proof}

\subsection{Proof of Proposition \ref{scrumSize}}
We first consider the number of binary comparisons given a preference over $X_n$. A binary comparison simply takes two alternatives $x,y \in X_n$ and ascribes either $x \succ y$ or $y \succ x$. As such, each preference has $\binom{n}{2}$ binary comparisons. By the definition of single-crossing, if $x \rhd y$ and $x \succ_i y$ then $x \succ_j y$ for all $j \geq i$. Let $\rhd^-$ denote the preference given by $x \rhd y \implies y \rhd^- x$. Thus the largest a set of single-crossing preferences could possibly be is the size of following construction. Let $\rhd^-$ be the first preference in our (ordered) set and let $\rhd$ be the last preference in our ordered set. Then given preference $\succ_i$, we simply change one binary comparison that agrees with $\rhd^-$ to one which agrees with $\rhd$. In terms of size, we start with a single preference and then iteratively change the ordering of each binary comparison. This corresponds to $\binom{n}{2}+1$ preferences.

    We have just constructed an upper bound for the size of a set of single-crossing preferences. We now give an algorithm that reaches that upper bound. Enumerate each alternative $x_i,x_j$ such that $i > j \implies i \rhd j$.
    \begin{enumerate}
        \item Initialize at $i=k=1$ and add $\rhd^-=\succ_k$ to the set $\mathcal{M}$.
        \item Set $j = i+1$.
        \item Take $\succ_k$ and construct $\succ_{k+1}$ by swapping the ranking of $x_i$ and $x_j$. Add $\succ_k$ to $\mathcal{M}$. Set $k=k+1$.
        \item Set $j=j+1$. If $j>|X|$, set $i=i+1$ and proceed to the next step. If not, return to the prior step.
        \item If $i=|X|$, terminate the algorithm. If not, return to step 2.
    \end{enumerate}
    Our algorithm effectively takes $\rhd^-$ as input, takes the lowest ranked alternative and iteratively moves it up one rank at each iteration until it is ranked in the same location as it is in $\rhd$. Since at each step of the algorithm, we are only changing the binary comparison of two alternatives which are ranked next to each other, in each step of the algorithm, we construct a preference. We start with $\rhd^-$ and then make $\sum_{i=1}^{n}\sum_{j=i+1}^{n}1$ swaps of binary comparisons. This corresponds to $\binom{n}{2}+1$ preferences. Thus we reach our upper bound from the first half of the proof and we are done.

\subsection{Proof of Proposition \ref{SCRUMEdge}}

\begin{proof}
    If $\nu$ has a single preference in its support, we are done. Suppose otherwise that $n\geq 2$, so that that $\succ_1$ and $\succ_2$ are in the support of $\nu$. Since $\succ_1$ and $\succ_2$ differ, this means that there exists some pair $(x,y)$ such that $x \succ_1 y$ and $y \succ_2 x$. Since $\nu$ satisfies the single crossing property, for all $i \geq 2$, we have that $y \succ_i x$. Let $A$ denote the set such that $\succ_1 \in L(x,A)$. Since $x \succ_1 y$, $y \in A$ and thus $L(x,A) \cap \{\succ_1,\dots,\succ_n\}=\{\succ_1\}$.  The result now continues by induction.
\end{proof}

\subsection{Proof of Proposition \ref{carumIdent}}
\begin{proof}
    The proof of edge decomposability of $M(\rhd)$ is discussed following Proposition \ref{carumIdent}. Now we consider the rest of the proposition. Suppose that there exists some $(\rhd,\nu)$ with $\nu \in M(\rhd)$ such that random choice rule $p$ satisfies $p=p_\nu$.
    \begin{cl}\label{priorclaim}
        For each $\emptyset\neq A \neq X_n$, there is at most one $x\in A$ such that $q(x,A) > 0$.
    \end{cl}
    \begin{proof}
        Consider two preferences $\succ$ and $\succ'$ with $\succ \in L(x,A)$ and $\succ' \in L(y,A)$ for $x \neq y$ and $A \neq X_n$. Without loss, we can take $\succ$ to be $\rhd$. In a Latin square, since $A \neq \emptyset$, there is a single preference which ranks $X \setminus A$ as the first $|X \setminus A|$ alternatives. In our case, this corresponds to $\rhd=\succ$. This means that $\succ' \not \in M(\rhd)$ and so we are done.
    \end{proof}
    There is some $\nu$ such that $p=p_\nu$. This means that each $q(x,A) \geq 0$ (by \citet{falmagne1978representation}) and that there is some $x \in X_n$ such that $q(x, X_n) > 0$. Now consider $X_n \setminus \{x\}$. By \citet{falmagne1978representation}, we have $\sum_{x \in A}q(x,A)= \sum_{y \in X_n \setminus A}q(y,A\cup \{y\})$ for each $A \neq X_n$. This means that there is some $y \in X_n \setminus \{x\}$ such that $q(y,X_n\setminus \{x\})>0$. By Claim \ref{priorclaim}, this $y$ is unique. We can repeat this process, repeatedly going from $A$ to $A \setminus \{z\}$ where $z$ satisfies $q(z,A)>0$. This process induces the unique path from $X_n$ to $\emptyset$ that passes through $X_n \setminus \{x\}$ and satisfies $q(y,A) >0$ along every edge of the path. This means that the preference corresponding to this path is in the support of $\nu$. Call this preference $\succ$. Since respects is an equivalence relation, any preference in the Latin square of $\succ$, the set of preferences which respect $\succ$, can be chosen to be $\rhd$. In this case, we can choose $\succ$ as $\rhd$, apply edge decomposability, and we are done.
\end{proof}

\bibliographystyle{ecta}
\bibliography{linrum}

\end{document}